\documentclass[12pt]{article}
\pdfoutput=1

\usepackage{amsmath}
\usepackage{amsfonts}
\usepackage{amssymb}
\usepackage{amsthm}
\usepackage{setspace}
\usepackage{color}
\usepackage{subcaption}
\usepackage{caption}
\usepackage{enumerate}
\usepackage[hidelinks]{hyperref}
\usepackage{graphicx}% Include figure files
\usepackage[hidelinks]{hyperref} %[linktocpage=true]
\usepackage[nottoc]{tocbibind}
\usepackage[utf8]{inputenc} 

\newcommand\cyan{\color{black}}
\newcommand\ee{\end{equation}}
\newcommand\be{\begin{equation}}
\newcommand\eea{\end{eqnarray}}
\newcommand\bea{\begin{eqnarray}}

\newcommand\comment[1]{}
\newcommand\expect[1]{\left\langle #1 \right\rangle}
\newcommand\bsb{\boldsymbol}
\comment{
\hypersetup{
colorlinks=true,
citecolor=DarkBlue,
linkcolor=DarkBlue,
urlcolor=DarkBlue,
}}

\def\O{\mathcal{O}}
\def\d{\partial}

\def\vep{\varepsilon}
\def\zrms{\zeta_{\rm rms}}
\def\Jrms{J_{\rm rms}}
\def\r{\bsb r}
\def\v{\bsb v}
\def\k{{\bsb k}}
\def\x{\bsb x}
\def\vphi{\varphi}
\def\vpbg{{\tilde\vphi}}
\def\J{\bsb J}
\def\ep{\epsilon}

\begin{document}

\begin{center}

  {\Large\bf Spin of Primordial Black Holes}

\vskip 1 cm
{\large Mehrdad Mirbabayi,$^{a,b}$ Andrei Gruzinov,$^{c}$ Jorge Nore\~na $^d$ }
\vskip 0.5 cm

{\em $^a$ International Centre for Theoretical Physics, Trieste, Italy}

{\em $^b$ Stanford Institute for Theoretical Physics, Stanford, CA, USA}

{\em $^c$ New York University, New York, NY, USA}

{\em $^d$ Instituto de F\'isica, Pontificia Universidad Cat\'olica de Valpara\'iso, Valpara\'iso, Chile}
\vskip 1cm

\end{center}
\noindent {\bf Abstract:} {\small Primordial black holes, formed from rare peaks in the primordial fluctuations $\zeta$, are non-rotating if the RMS (root mean square) fluctuation $\zeta_{\rm rms}$ is sent to zero. We show that the spin also vanishes at first order in $\zrms$, suggesting the dimensionless spin parameter $a_{\rm rms} \sim \zrms^2$. We identify one quadratic contribution to the spin by calculating (and extrapolating to the formation time) the torque on a black hole due to ambient acoustic waves. For a reasonable density of primordial black holes this implies a percent level spin parameter.}

\vskip 1 cm

\section{Introduction}

It is an intriguing idea that primordial black holes (PBHs) make up a significant fraction of dark matter \cite{Chapline}. This possibility is seriously challenged by various cosmological and astrophysical observations \cite{Carr,Seljak,Katz,Byrnes}. Nevertheless, further investigation is underway and it is worthwhile having a clear understanding of PBH properties. Properties of particular interest are PBH mass distribution, clustering, and spin distribution given an underlying formation scenario.

Several such scenarios have been proposed. A common theme is to invoke an enhancement of the almost Gaussian primordial spectrum of fluctuations $\zeta$, from $\zrms \sim 10^{-5}$ to $\zrms \sim 0.1$. To form PBHs of average mass $\bar M$ the enhancement is over a range of scales around 
\be\label{k0}
k_0 \sim 10^9 \sqrt{\frac{M_\odot}{\bar M}} k_{\rm eq},
\ee
where $k_{\rm eq}$ is the matter-radiation equality momentum scale. See for instance \cite{Motohashi,Clesse} for implementations in the inflationary context.

Black holes form when rare peaks with height $\zeta_0 > \zeta_c \sim 1$ in the initial random field enter the horizon. In a conventional thermal history, this happens during radiation-dominance when the scale factor grows as $t^{1/2}$, explaining the above formula for $k_0$ (the horizon mass at matter-radiation equality is about $10^{18}M_\odot$). 

The need for an enhancement of $\zrms$ is a consequence of the exponential suppression of the number density of high peaks. For a Gaussian spectrum, the energy density in PBHs at the formation time is $\rho_{\rm PBH}\sim \rho_{\rm tot}\exp(-\zeta_c^2/2\zrms^2) $. By the time of matter-radiation equality this ratio grows by the ratio of scale factors giving
\be
\frac{\Omega_{\rm PBH}}{\Omega_m} \sim  10^9 \sqrt{\frac{M_\odot}{\bar M}} e^{-{\zeta_c^2}/{2\zrms^2}} 
\ee
explaining the estimate $\zrms \sim 0.1$ for an order-one threshold $\zeta_c$.

The precise value of the threshold for PBH formation and its shape-dependence have been extensively studied in the literature starting from the pioneering works of Carr and Hawking \cite{Carr_Hawking,Carr_mass}. It is often formulated in terms of a critical value for $\delta\rho/\rho$ (after fixing an appropriate gauge, such as the comoving gauge). This naturally eliminates irrelevant long-wavelength fluctuations of $\zeta$. Nevertheless for a fixed shape of the peak with characteristic size $1/k_0$, it is useful to have a criterion in terms of $\zeta$ which is nearly Gaussian in single-field models of inflation. Such a criterion can be found in \cite{Musco} and it is implemented in the derivation of PBH abundance in \cite{Yoo}. 

All peaks of height $\zeta_0>\zeta_c$ collapse into black holes, however as a remarkable example of Choptuik scaling \cite{Choptuik}, the mass of the resulting black holes follow a scaling relation
\be
M(\zeta_0) = K \bar M (\zeta_0-\zeta_c)^\gamma,
\ee
with $K=\O(1)$ and $\gamma \simeq 0.39$ as long as $\zeta_0$ is not too much larger than $\zeta_c$ \cite{Niemeyer,Musco}. Hence even though $\bar M$ is of the order of the horizon mass corresponding to the horizon crossing time of $k_0$ (as already implied in \eqref{k0}), the initial mass function for PBHs extends all the way to $M=0$ \cite{Niemeyer_IMF}.

One may question the validity of the Gaussian approximation for $\zeta$ in this setup, given that PBHs are sensitive to the tails of the distribution. Suppose non-Gaussianities are parametrized (schematically) as $\zeta = g + f_{\rm NL} g^2 +\O(g^3)$, where $g$ is Gaussian and the nonlinear relation can in principle be non-local. If $f_{\rm NL} \zeta_c \sim 1$ one would expect non-Gaussianities to be important. And this is compatible with the current CMB constraints. However, the approximation is justified in a minimal slow-roll model since there $f_{\rm NL} \ll 1$ \cite{Maldacena}. 

Furthermore, in the absence of local-type non-Gaussianity (as in the conventional single-field models of inflation) the effect of long-wavelength fluctuations $\zeta_{\k}$ on any local physical process, including PBH formation, is suppressed by $k^2/k_0^2$ \cite{Creminelli}. Therefore there is no significant clustering of PBHs. At long wavelengths $\delta\rho_{\rm PBH}/\bar\rho_{\rm PBH}$ follows the primordial field $\zeta$, as does any thermal relic density \cite{Ali-Haimoud}.\footnote{To be clear, both the slow-roll approximation and the resulting insignificance of non-Gaussianities can be broken in specific models that predict abundant PBH production. See for instance \cite{Motohashi,Germani,Franciolini,Atal}. Nevertheless since there is no unique way to be non-Gaussian, in this paper we focus on Gaussian models as a benchmark.}

\begin{figure}[t]
\centering
\includegraphics[scale =0.7]{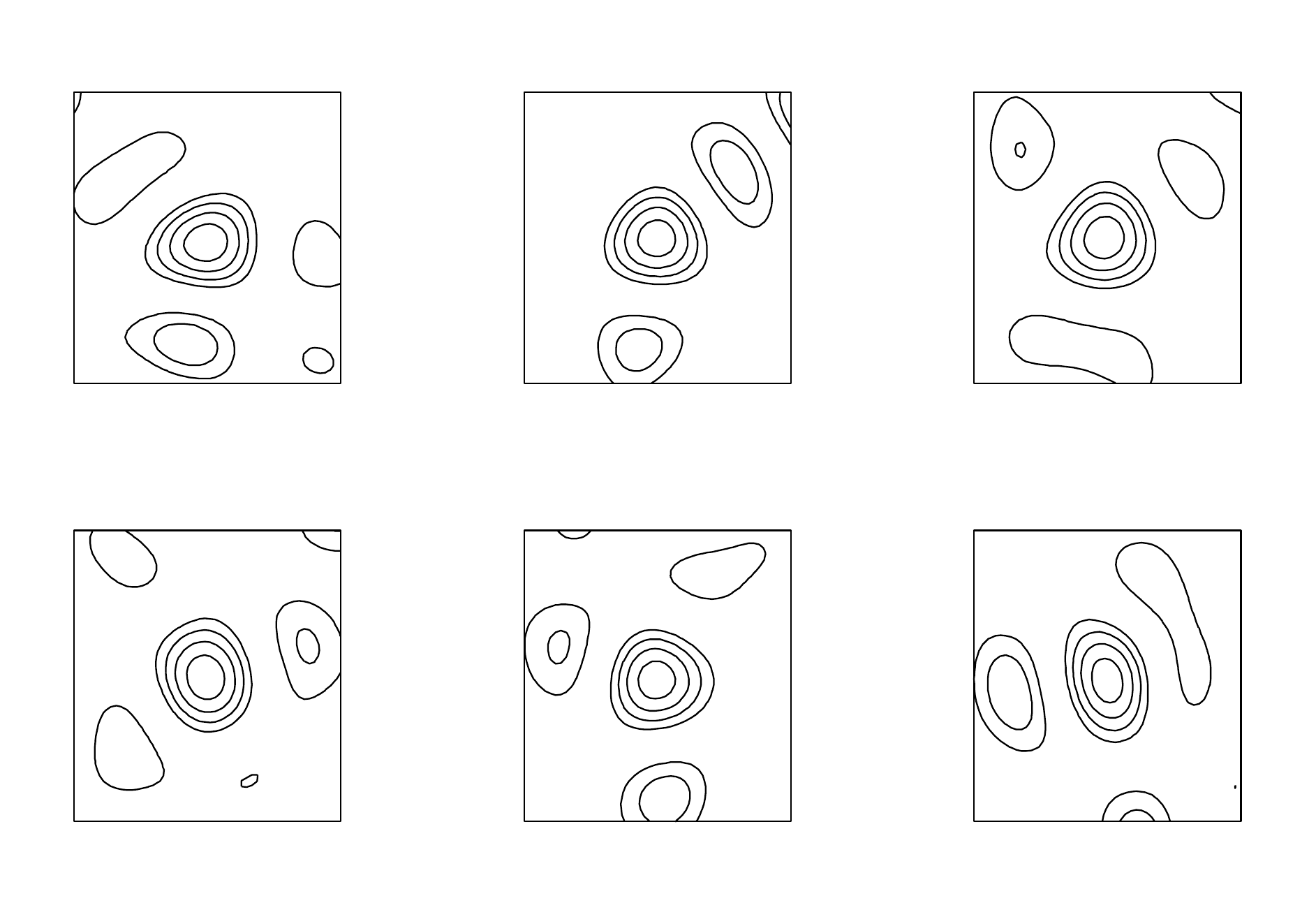} %[width= 11cm,height=8cm]                                                      
\caption{\small{\cyan A mosaic of six peaks higher than 5.6 sigma in a two-dimensional Gaussian random field. The levels are 2,3,4,5 sigma and the power spectrum is $P(k)=k^4 \exp(-k^2)$.}}
\label{fig:peaks}
\end{figure}

Our goal here is to pursue a similar level of analytic insight into the problem of spin distribution of PBHs (which seems to be missing despite an earlier work \cite{Chiba}). The spin is conventionally believed to be small. The reason behind it is the approximate spherical symmetry of rare peaks in a Gaussian random field. Quantitatively, the extra symmetry means that the eigenvalues of the Hessian matrix at the peak, which we denote as $(\d_1^2\zeta_0,\d_2^2\zeta_0,\d_3^2 \zeta_0)$, are close in a relative sense \cite{BBKS}. {\cyan For instance, for a spectrum dominated by a single scale $k_0$,}
\be\label{bbks}
\frac{\d_1^2 \zeta_0 -\d_2^2 \zeta_0}{\nabla^2 \zeta_0} \sim \frac{\zrms}{\zeta_0}.
\ee
This is what one would expect if typical fluctuations were superposed on a perfectly spherical peak of height $\zeta_0$ (see figure \ref{fig:peaks}). Given that $\zrms \sim 0.1$ and $\zeta_0\sim 1$ in our problem, the key question is at what order these fluctuations contribute to the spin $J_{\rm rms}$.\footnote{Deviation from sphericity can also affect the collapse threshold as discussed in \cite{Kuhnel}.}

%%%%%%%%%%%%%%%%%%%%%%%%%%%%%%%%%%%%%%%%%%%%%%%%                                              
%%%%%%%%%%%%%%%%%%%%%%%%%%%%%%%%%%%%%%%%%%%%%%%           

An educated guess, and what we are going to argue for, is that for mass-$M$ black holes
\be\label{Jrms}
\Jrms \sim M r_g \zrms^2,
\ee
where $r_g = 2G M$ is the gravitational radius. This guess is motivated by Peebles' analysis of the origin of the angular momentum of galaxies \cite{Peebles}, where the total angular momentum contained in a spherical  proto-galaxy is shown to start at second order in perturbation theory. However Peebles' argument is Newtonian and does not apply directly to PBHs, which form at cosmological scales. There is no general relativistic notion of angular momentum at cosmological scales, even though it makes sense to talk about the spin once the cosmological horizon grows much larger than their size. 

In section \ref{sec:first} we will show that by symmetry the relativistic expression for $J$ vanishes at first order in deviations from sphericity. This supports the estimate \eqref{Jrms} by showing that there is no $\O(\zrms)$ contribution. {\cyan We also show that the second order contribution (if present) has to be suppressed by $\Delta k/k_0$ if the range of enhanced modes is narrow, $\Delta k\ll k_0$. However such a narrow spectrum is neither required nor obviously viable.}

As an evidence that the second order contribution is non-vanishing, in section \ref{sec:second} we will calculate the torque on a black hole from ambient acoustic waves at the time $t$ much later than the BH formation time $\sim r_g$. This creates a hierarchy between the black hole size $\sim r_g$, the characteristic wavelength of the acoustic waves in the background radiation fluid $\frac{a(t)}{k_0}$, and the cosmological horizon $1/H=2t$:
\be\label{hierarchy}
r_g\ll \frac{a(t)}{k_0} \ll t.
\ee
Hence one can calculate the instantaneous torque due to a perturbed Bondi accretion at leading order in gradient expansion:
\be\label{tau}
\bsb{\dot J} =\frac{1}{3}\lambda^2 \dot M r_g^2 \ \nabla\delta \times \bsb v,
\ee
where $\lambda \simeq 0.867$ and 
\be
\dot M =8\sqrt{3}\pi r_g^2 \bar\rho
\ee
is the rate of mass accretion onto a black hole immersed in an asymptotically uniform radiation fluid of density $\bar\rho$. The density contrast $\delta =\delta\rho/\bar\rho$ and the fluid velocity $\bsb v$ are related (in the cosmological setting) to the primordial field $\zeta$. 

We will calculate $\Delta\Jrms(t)$, the rms value of total angular momentum gained after time $t$. By definition $\Delta \Jrms(t)$ monotonically increases toward earlier $t$, and significantly so due to the cosmic dilution. {\cyan Assuming a Gaussian shell of enhanced modes around $k=k_0$ and with width $\Delta k<k_0$, we obtain
\be\label{DJ0}
\Delta J_{rms}(t)\simeq M r_g \zrms^2 \frac{\Delta k}{k_0} \frac{{3}\lambda^2 H_0^2r_g^2}{(k_0/aH)^2}.
%\left(\frac{\sqrt{3}H_0r_g}{(k_0/aH)}\right)^2
\ee
Here $H_0$ is the Hubble rate at the time when $k_0$ crosses the horizon. A characteristic-mass PBH has $H_0 r_g \sim 1$. Extrapolating \eqref{DJ0} to the horizon-crossing time $k_0/aH\sim 1 $ (when our approximation breaks down) and taking the width $\Delta k$ not so much less than $k_0$} predicts an order-one coefficient in \eqref{Jrms}, barring cancellation with other contributions to $J$ from the black hole formation process. We conclude in section \ref{sec:con}.

%%%%%%%%%%%%%%%%%%%%%%%%%%%%%%%%
\section{Perturbation theory around spherical collapse }\label{sec:first}

The key conceptual insight into the problem of PBH spin is the above-mentioned smallness of deviations from sphericity. A high peak in the primordial fluctuations that collapses into a black hole can be decomposed into a collapsing spherically symmetric component plus order $\zrms$ fluctuations. The mass, position, momentum and angular momentum of the resulting black hole have a perturbative expansion in those fluctuations. 

This decomposition has $\O(\zrms)$ arbitrariness that leads to $\O(\zrms)$ ambiguities in the zeroth order mass $M$ and the zeroth order position of the black hole (which will be taken as the origin $\r =0$). The ambiguities are systematically cancelled as one proceeds in the perturbative expansion. On the other hand, the momentum and angular momentum are unambiguously zero in the spherically symmetrical approximation. In practice going beyond zeroth order in this perturbative expansion around the $t$- and $r$-dependent collapsing background is a daunting task. Our strategy is to use the formulation of perturbative expansion to show that the first order spin $J_{(1)}=0$ on symmetry grounds.

Another feature of the PBH problem that needs extra care is the breakdown of Newtonian approximation to Einstein gravity. As a result, mass, momentum and angular momentum cannot be localized. Strictly speaking they are defined only asymptotically for certain asymptotic geometries such as in asymptotically flat spacetimes. In practice it is of course sensible to talk about PBH parameters at late enough times (certainly today). A necessary condition is that the size of the black hole $r_g$ be much smaller than the cosmic time $t$. We will see that it is also sufficient, as long as we care about spin up to $\O(\zrms^2)$. 

To talk about the black hole parameters one first needs to define (and even to be able to define) an approximately inertial frame around the black hole. The ``size of the frame'' $R$ has to be large enough such that at distance $R$ the deviation from Minkowski metric due to the black hole curvature is small, namely $r_g/R\ll 1$. Black hole parameters can then be unambiguously defined in terms of the properties of Keplerian orbits at $r\sim R$. However there is another source of deviation from Minkowski caused by the cosmological curvature. This is of order $H^2 R^2=R^2/4t^2$. So, for instance, to measure mass accurately we need this to be much smaller than the black hole field $r_g/R$:
\be
R^3\ll r_g t^2.
\ee
To measure the spin, there is another more stringent upper bound on $R$. This is because the radiation fluid that surrounds our black hole also carries angular momentum in the presence of fluctuations. For a given precision $\ep$ in black hole spin one needs to be able to choose $R$ to be small enough such that the total angular momentum of the radiation fluid inside the region $r_g<r<R$ is less than $\ep$. This criterion is sensible despite the ambiguity of the lower limit because angular momentum is dominated by large $r\sim R\gg r_g$. Hence we can use the flat space expression 
\be\label{Jrad}
J_{\rm rad}^{i} \simeq\frac{4}{3} \vep^{ijk} \int_{r<R} d^3 \r  \ \rho r^j v^k,
\ee
where $\vep^{ijk}$ is the fully anti-symmetric tensor and repeated indices are summed over.\footnote{The (perhaps unfamiliar) factor of $4/3$ is because we are dealing with a relativistic fluid with pressure $p=\rho/3$. Stress-energy and angular momentum conservation of radiation fluid will be discussed at length in section \ref{sec:second}.} To leading order in cosmological perturbations $v^i$ is a pure gradient $v^i = \d_i \tilde\vphi$. Hence to first order the above integral reduces to a surface term
\be\label{Jrad1}
J_{{\rm rad}(1)}^i(r<R)\simeq \frac{4}{3}\vep^{ijk} \bar\rho R^2 \int_{r=R} d^2\hat r \ \tilde\vphi\ r^j \hat r^k =0,
\ee
where we used the fact that the normal to the spherical surface $r=R$ is $\hat r$. The above argument is essentially the same as Peebles' for the vanishing of the first order angular momentum of proto-galaxies. 

At second order $J_{\rm rad}$ is non-vanishing. It can be estimated by taking into account the spatial variations in density $\rho$. (As we will see the vorticity in radiation fluid is $\nabla\times \v =-\frac{1}{4\rho}\nabla\rho \times\v$ giving the same estimate.) In the tightly coupled regime the amplitude of the acoustic waves remains constant and of order $\zrms$. This is reviewed below in section \ref{sec:cospert}. Their typical wavelength $a(t)/k_0$ has by now significantly stretched and can be assumed to be longer than $R$. We therefore replace $\rho \to  \r\cdot \nabla\rho  \sim \bar \rho r (k_0/a(t)) \zrms $ and $v\to \zrms$, and obtain
\be
J_{\rm rad} (r<R) \sim \bar \rho \frac{k_0}{a(t)}R^5 \zrms^2.
\ee
The scale factor grows as $a(t) \propto t^{1/2}$, so we can write $\bar \rho \sim M/(r_g t^2)$ and $k_0/a(t) \sim 1/\sqrt{t r_g}$, giving finally
\be
J_{\rm rad}(r<R) \sim M r_g \zrms^2 \left(\frac{R^2}{r_g t}\right)^{5/2}.
\ee
Hence if the black hole has a nonzero spin at or below second order in $\zrms$, the angular momentum inside the region $r<R$ is guaranteed to be dominated by the hole itself if
\be\label{hier}
r_g\ll R\ll \sqrt{r_g t}.
\ee
For such a choice to be available it is enough to have $t\gg r_g$, as anticipated.\footnote{To discuss black hole spin at $n>2$nd order one needs to wait parametrically longer, $t\gg r_g \zrms^{-2(n-2)/5}$.}

The general relativistic expression for the total angular momentum in the region $r<R$ is given for instance in \cite{Landau} in terms of $\tau^{\mu\nu}$, the pseudo-tensor of energy and momentum,
\be\label{J}
J^{i}= \vep^{ijk}\int_{r<R} d^3\r \ r^j\tau^{0k}.
\ee
Note that this is a vector defined by observers in the near Minkowski asymptotic region. So we can liberally raise and lower spatial indices and sum over repeated ones (our choice of metric signature is mostly plus). The coordinate system is uniquely fixed at large radii $r\gg r_g$ by the requirement that it is an inertial frame, and at rest with respect to the asymptotic FRW cosmology. On the other hand the extension to the interior (possibly inside the black hole event horizon) is completely arbitrary. Nevertheless $J$ is well-defined because using the Einstein equations the pseudo-tensor is related to the linearized Einstein tensor,
\be
\tau^{\mu\nu}=\frac{1}{8\pi G} G^{\mu\nu}_{\rm lin};
\ee 
$G^{\mu\nu}_{\rm lin}$ is a total derivative and the integral reduces to a surface term at $R$; the surface term is invariant under linearized diffeomorphisms.\footnote{The definition of angular momentum in general relativity is somewhat subtler than energy and momentum, even in asymptotically flat spacetimes. Physically this is because a graviton with arbitrarily small energy can carry away a finite amount of angular momentum, and correspondingly there are metric configurations with this property \cite{Huang,Chen}. In the realistic situation, long after the initial formation period there always exists a frame in which linearization in $g_{\mu\nu}-\eta_{\mu\nu}$ is valid at large radii $r\sim R$ and the expression \eqref{J} is unambiguous.}

We now use the perturbation theory to calculate $J$. At zeroth order, a perfectly spherical over-density collapses into a Schwarzschild black hole at $\r =0$. Thus $J$ has to vanish. Indeed at zeroth order in $\zrms$, the only possible form of $\tau^{0k}$ is 
\be
\tau_{(0)}^{0k} = f_0(t,r) \hat r^k,
\ee
for some function $f_0$. Substitution in \eqref{J} gives zero. At first order in perturbations there are two possible structures
\be\label{tau1}
\tau_{(1)}^{0k} = f_{1}(t,\r) \hat r^k + \d_k g_1(t,\r),
\ee
where now $f_1$ and $g_1$ are two scalar fields, linearly related to the initial perturbations.\footnote{Exterior products are excluded because the momentum density $\tau^{0k}$ has to be parity odd.} Both terms give vanishing contributions to the angular momentum \eqref{J}: the $f_1$ term for the same reason as the zeroth order contribution, and the $g_1$ term reduces to a boundary integral 
\be
\vep^{ijk} \int_{r=R} d^2\hat r \ g_1 r^j \hat r^k =0.
\ee
Note that the scalar nature of primordial fluctuations was crucial, otherwise other tensorial structures could appear in \eqref{tau1}. 

To complete the argument we note that even though there can be a first order displacement $\r_{(1)}=\O(\zrms)$, there is no first order center of mass angular momentum (which should have been subtracted from the total angular momentum to obtain the spin). This is because black hole momentum vanishes at zeroth order:
\be
P^i_{(0)} = \int_{r<R} d^3\r \ \tau_{(0)}^{0i}=0.
\ee
Hence $J$ can at best be nonzero at $\O(\zrms^2)$. 

{\cyan One last observation to make on purely symmetry grounds: any second order contribution to $J$ has to rely on combining modes $\zeta_\k$ in the initial spectrum with unequal $k=|\k|$. If the initial power spectrum is dominated by an infinitely narrow shell of momenta then $J_{(2)}=0$. This is because the most general second order expression is
\be
\bsb J_{(2)}=\int_{\k_1,\k_2} \zeta_{\k_1}\zeta_{\k_2}  F(k_1,k_2) \ \k_1\times \k_2
\ee
where $\int_\k = \int \frac{d^3\k}{(2\pi)^3}$ and $F(k_1,k_2)$ is an anti-symmetric kernel. By anti-symmetry $F(k_1,k_2)$ vanishes at zeroth order in $k_1 - k_2$. }

To summarize, we have argued that the spin of PBHs is a second order quantity. The essence of the argument is that at linear order one cannot construct a pseudo-vector (such as angular momentum) out of a scalar field. The means to this end was writing the relativistic expression for the total angular momentum inside a spherical region of radius $R$ that satisfies \eqref{hier}. The choice of a spherical region is harmless because, as argued above, the angular momentum in the region is dominated by the black hole and therefore one can deform the boundary of the region to make it spherically symmetric. For a nearly Gaussian primordial field, equation \eqref{bbks} then implies that $\Jrms\propto \zrms^2$.\footnote{One often encounters the contrary claim that angular momentum is a first-order quantity, both in the literature on structure formation such as \cite{Doroshkevich,White} and perhaps motivated by that also in the PBH context \cite{Riotto}. Our symmetry argument applies to any collapsed structure formed from rare (and hence nearly spherical) peaks of a scalar primordial field. It rules out the claim that $J$ starts at first order in perturbations. The claim seems to originate from the unjustified assumption that collapsing regions have a mass distribution with order-1 anisotropy that is acted upon by a first order external torque. However, the shapes of the collapsing regions are not arbitrary. Conditioned on having a peak of a certain height, and sending the rms fluctuations to zero, one obtains a perfectly spherical peak as implied by \eqref{bbks}. So the anisotropy of the collapsing region is itself a first order quantity. We are thankful to Marcello Musso and Ravi Sheth for a useful discussion on this point.}

%%%%%%%%%%%%%%%%%%%%%%%%%%%%%%%%%%%%%%%%%%%%%%%%%%
\section{Acoustic spin up of primordial black holes}\label{sec:second}

We are not able to calculate the spin at second order. In the absence of a symmetry reason for it to vanish, we expect $\Jrms$ not to be much less than $M r_g \zrms^2$ {\cyan unless $\Delta k\ll k_0$}. To support this expectation we calculate $\Delta \Jrms (t)$, the rms change of angular momentum after time $t\gg r_g$. This is calculable because by this time there is a clear separation of scales. The size of the black hole is much less than the characteristic wavelength of the acoustic waves, and the latter much shorter than the horizon size. The surrounding neighborhood of the black hole has been randomized by the propagation of the sound waves, and finally, black hole peculiar velocity (an $\O(\zrms)$ memory of the formation time) has red-shifted away. In such a circumstance, we expect a nearly spherical Bondi accretion, that is perturbed by the long-wavelength acoustic waves, and that therefore exerts a torque on the black hole. All these approximations break down as $t\to r_g$, but the extrapolation does give an order one coefficient in the formula $\Jrms\sim M r_g \zrms^2$.\footnote{This extrapolation scheme is partly inspired by Peebles' work \cite{Peebles}.}

Another simplification is that the radiation fluid has a purely potential flow since there is no vector perturbations in the super-horizon initial conditions. This implies that in the dissipation-less regime the system is fully described by a scalar field $\vphi$, coupled to Einstein-Hilbert gravity \cite{Boubekeur}:
\be
S=S_{\rm EH} +\int d^4x \sqrt{-g} p(X),\qquad X \equiv -g^{\mu\nu} \d_\mu \vphi \d_\nu \vphi.
\ee
The function $p(X)$ is fixed by the requirement that the stress-energy tensor of the scalar field action
\be\label{stress}
T_{\mu\nu} = 2 p'(X) u_\mu u_\nu + g_{\mu\nu} p(X),\qquad u_\mu \equiv \frac{\d_\mu\vphi}{\sqrt{X}},
\ee
matches that of a perfect fluid with sound speed $c_s^2=1/3$. This fixes $\rho(X) = 3 p(X) = X^2$. Therefore the fluid equations reduce to
\be
\d_\mu (\sqrt{-g} g^{\mu\nu} X \d_\nu \vphi)=0.
\ee
%Hence the combination $\rho^{1/4} u_\mu$ is curl-free, which for non-relativistic velocities implies the velocity-vorticity is not an independent field:
%\be
%\nabla\times \v = -\frac{1}{4} \frac{\nabla \rho}{\rho} \times \v.
%\ee
In the neighborhood of a PBH and for large enough $t$ such that we can take $r\ll t$, the fluid self-gravity is negligible and the metric is well approximated by Schwarzschild
\be
ds^2 = -\left(1-\frac{r_g}{r}\right)dt^2 + \frac{dr^2}{\left(1-\frac{r_g}{r}\right)} +r^2 (d\theta^2 +\sin^2\theta d\phi^2).
\ee
On this background the radiation fluid accretes onto the black hole. If the fluid is perturbed, the accreting fluid also carries momentum and angular momentum into the black hole. Below we will first calculate the accretion rate and the resulting torque. Afterward we will calculate $\Delta \Jrms(t)$ in the cosmological scenario described in the Introduction.

%%%%%%%%%%%%%%%%%%%%%%%%%%
\subsection{Unperturbed Bondi flow}

Finding the steady-state mass accretion rate $\dot M$, at zeroth order in perturbations, is similar to the standard Bondi problem \cite{Bondi}. One seeks a spherically symmetric solution for $\vphi$ that matches the background solution at large $r$ and is regular at the horizon. Far from the black hole (but well inside the cosmological horizon) the fluid is approximately at rest and uniform
\be
\dot{\bar\vphi} =-\bar\rho^{1/4} ={\rm constant}.
\ee
(The slow variation of $\bar\rho$ on cosmological time-scales will be trivially included when needed in order to calculate $\Delta \Jrms(t)$.) 

A stationary flow is described by $\vphi = \bar\rho^{1/4}(-t+t_0+\vphi_0(r))$, where $t_0$ is an irrelevant integration constant and $\vphi_0$ satisfies
\be\label{phi0}
(r^2 (1 - (1-r_g/r)^2 {\vphi_0'}^2) \vphi_0')'=0,
\ee
and prime denotes $d/dr$. Integrating once gives
\be\label{cube} 
u-u^3=c(r_g^2/r^2-r_g^3/r^3),
\ee
\be\label{u}
u\equiv-(1-r_g/r)\vphi_0',
\ee
with some constant $c$. Here $u$ is the infall physical velocity, that is, the velocity measured by a locally inertial observer who is instantaneously at rest at radius $r$. We must have
\be\label{asymu} 
u\rightarrow 0,~~r\rightarrow \infty;~~~~~u\rightarrow 1,~~r\rightarrow r_g.
\ee
Indeed, from \eqref{cube} we have $u\rightarrow \pm 1$ or $0$ at $r\rightarrow r_g$, but $u\rightarrow 0$ is actually singular at the horizon, with infinite pressure, while $u\rightarrow -1$ describes an excreting rather than accreting BH. 

%For small values of $c$ (and thus $\dot M$), the decaying solution approaches $0$ at the horizon, while the $u(r_g)= 1$ has wrong asymptotic behavior (see figure \ref{fig:bondi}). As $\dot M$ is increased there is a unique value at which the two solutions cross at a radius $r_s$, called the sonic point. At this value of $\dot M$ the asymptotically decaying solution connects to the regular free-falling solution at the horizon. For larger $\dot M$ there is no stationary inflow for an intermediate range of radii.}

%%%%%%%%%%%%%%%%%%%%%%%%%%%%%%%%%%%%%%%%%%%%%%%%                                              
\begin{figure}[t]
\centering
\includegraphics[scale =0.7]{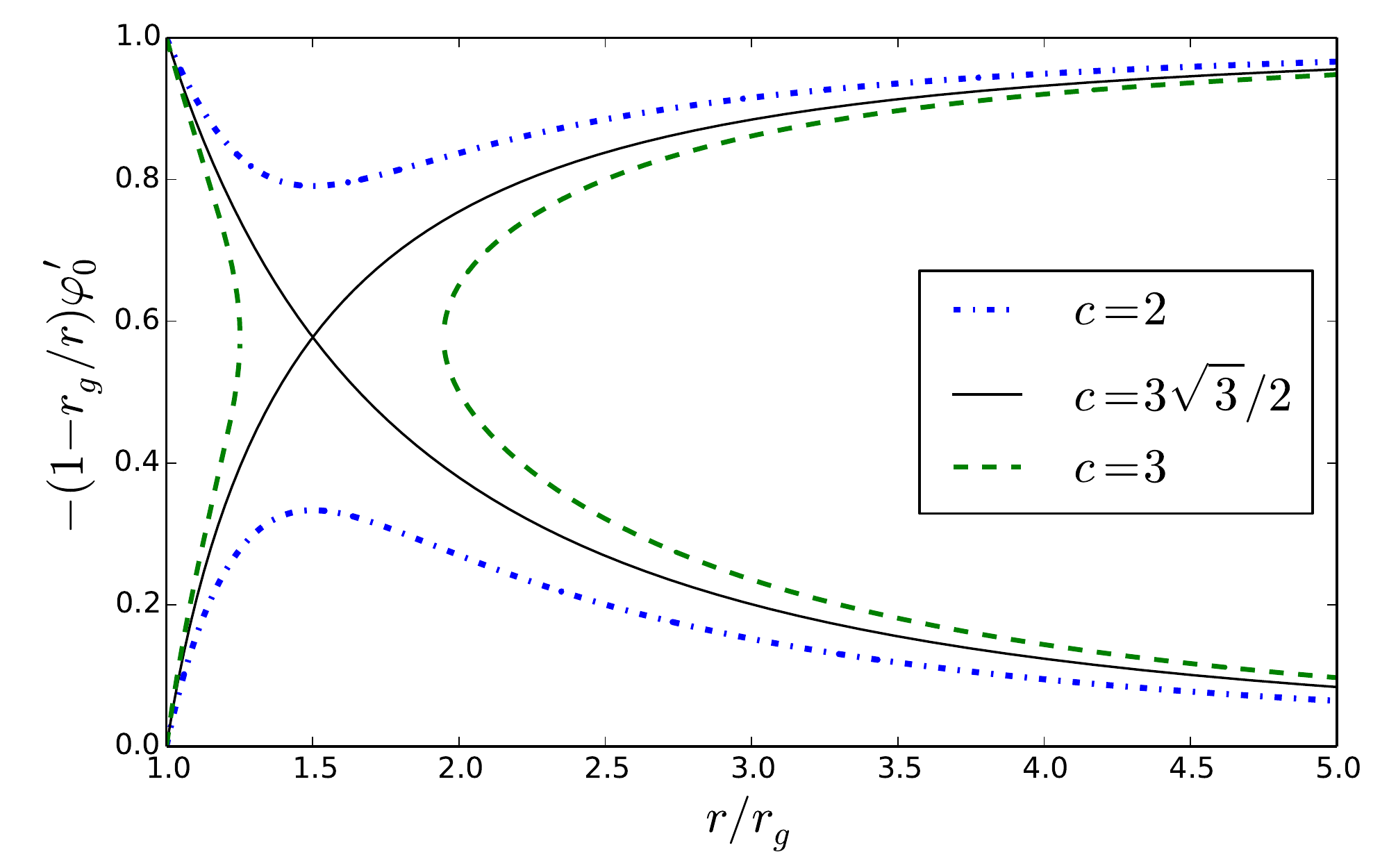} %[width= 11cm,height=8cm]                                                      
\caption{\small{\cyan Infall fluid velocity $u$ as a function of $r$. At low accretion rate, corresponding to $c<3\sqrt{3}/2$, the decaying solution at large $r$ is singular at horizon, while the regular solution has wrong asymptotic behavior. As $\dot M$ is increased, at $c=3\sqrt{3}/2$ the two solutions cross at the sonic radius, and the solution with correct asymptotics connects to the regular solution at the horizon. For larger $\dot M$ there is no steady-state inflow for a range of radii.}}
\label{fig:bondi}
\end{figure}
%%%%%%%%%%%%%%%%%%%%%%%%%%%%%%%%%%%%%%%%%%%%%%%           

Just like in the standard Bondi problem, the desired asymptotic behavior, \eqref{asymu}, is possible {\it iff} the l.h.s. and the r.h.s. of \eqref{cube} have the same maximal value. The maximal value of the l.h.s. is $\frac{2}{3\sqrt{3}}$ at the sonic point $u=\frac{1}{\sqrt{3}}$. The maximal value of the r.h.s. is $\frac{4}{27}c$ at the sonic radius $r_s=\frac{3}{2}r_g$, giving a unique value for the integration constant $c=\frac{3\sqrt{3}}{2}$ and a unique accretion flow 
\be\label{cube1} 
u-u^3=\frac{3\sqrt{3}}{2}(r_g^2/r^2-r_g^3/r^3),
\ee
where the root of the cubic is selected so as to be continuous and with correct asymptotics \eqref{asymu}, see figure \ref{fig:bondi}. 

Having fixed the solution for $u$ and $\vphi$ one calculates $\dot M$ as follows. Recall that in curved spacetime whenever the metric has a symmetry, i.e. a Killing vector field $\xi$, there is a conservation law:
\be
\d_\mu(\sqrt{-g} T^\mu_\nu \xi^\nu) =\sqrt{-g} \nabla_\mu(T^\mu_\nu \xi^\nu)=0.
\ee
The second equality follows from the covariant conservation of stress-energy tensor, plus the defining property of Killing vector fields $\nabla_\mu \xi_\nu=-\nabla_\nu \xi_\mu$. In our setup the Schwarzschild metric is time-independent, so $\xi_t = \d_t$ is a Killing vector field. The associated conservation law is energy conservation:
\be\label{Eflux}
\d_\mu (r^2 T^\mu_t) =0.
\ee
On the stationary solution the time-derivative vanishes, and this equation describes a constant flux of energy. Thus we can calculate the flux at any radius (using $\dot M = - \dot P_t$ in our convention)
\be
\dot M =  \int d^2\hat r r^2 T^r_t.
\ee
At large radii we have $T^r_t =-\frac{4}{3}\bar\rho \vphi_0'$. Using $\vphi_0'(r\gg r_g)=-c r_g^2/r^2$ gives
\be
\dot M = 8\sqrt{3} \pi r_g^2 \bar\rho.
\ee

%%%%%%%%%%%%%%%%%%%%%%
\subsection{Perturbed Bondi flow}

The Bondi flow is time-independent and spherically symmetric, so perturbations can be organized by frequencies and spherical harmonics.\footnote{The explicitly time-dependent term $\bar\vphi=\bar \rho^{1/4} (t_0-t)$ does not lead to any time-dependence in the equations of motion. It is only $\d_\mu\vphi$ and not $\vphi$ itself that matters.} We write
\be\label{vphi1}
\vphi=\bar\rho^{1/4}(-t+t_0 +\vphi_0(r)+\vphi_1(t,\r))
\ee
where
\be\label{phiomega}
\vphi_1(t,\r) = \sum_{l,m} \int \frac{d\omega}{2\pi} \vphi_{l,m}(\omega,r)  Y_{lm}(\hat r) e^{-i\omega t}.
\ee
At $r\gg r_g$ the perturbations have to match the background acoustic waves which can be expanded in plane-waves
\be
\label{asym}
\vpbg(\omega,\r) 
= \int d^2\hat k \vpbg(\omega,\hat k) e^{i \k\cdot \r},\qquad \k = \frac{\omega}{c_s}\hat k.
\ee
The characteristic frequency in the PBH problem is $\omega\sim (r_g t)^{-1/2} \ll 1/r_g$. Hence there is a region $r_g\ll r\ll 1/\omega$ where one can expand \eqref{asym} in gradients (i.e. powers of $\omega r$) and match to \eqref{phiomega}. The $l$-th harmonic is seen to start at $l$-th order in gradients. Below we will use this expansion to derive the leading contribution in $\omega r_g$ to the torque. Since torque is a pseudo-vector the leading contribution comes from combining two gradients, i.e. two $l=1$ modes, in an anti-symmetric way. 

Let's focus on the flux of $J^z$ into the black hole. The Schwarzschild metric is spherically symmetric, and in particular, it has a Killing vector field $\xi_\phi = \d_\phi$. The associated conservation law is $J^z$ conservation:\footnote{Note in passing that in flat spacetime and when the fluid velocity is non-relativistic, the $J^z$-density in \eqref{Jflux} is $r^2T^t_\phi=\frac{4}{3}r^2 X^{3/2} \d_\phi\vphi+\O(v^3)$. In Cartesian coordinates this reproduces eq. \eqref{Jrad}.}
\be\label{Jflux}
\d_\mu (r^2 T^\mu_\phi) =0.
\ee
When $\omega r_g\ll 1$ we can approximate the flow as stationary for 
\be
 r\ll  r_\omega\equiv \left(\frac{r_g^2}{\omega}\right)^{1/3},
\ee
namely over the region where the Bondi flow arrives at the horizon before a characteristic oscillation time $1/\omega$. In this region the time-derivative in \eqref{Jflux} is negligible and the equation implies that we can calculate the rate of angular-momentum accretion onto black hole (i.e. the torque) at any radius:
\be
\dot J^z=- \int d^2\hat r r^2 T^r_\phi.
\ee
Substituting \eqref{vphi1} in the expression for the stress-energy tensor \eqref{stress}, we find, with no surprise, that the first nonzero contribution to the torque appears at second order in $\vphi_1$
\be\label{tauz}
\dot J^z =- \frac{4}{3}\bar \rho \int d^2\hat r r^2[(1 - 3 (1-r_g/r)^2 {\vphi_0'}^2) \vphi_1'-2 \vphi_0' \dot \vphi_1]
\d_\phi\vphi_1.
\ee
Note that the two $\vphi_1$ perturbations have to have the same $l$ and opposite and nonzero $m$ for the torque to be nonzero. This confirms our expectation that the leading contribution in gradient expansion comes from two $l=1$ modes. 

It is also easily verified that the integrand reduces to a total $\phi$-derivative when both $\vphi_1$'s have the same frequency. In fact, there is a clever choice of radius $r=r_s$ which makes this point manifest. At the sonic point
\be\label{sonic}
1 - 3 (1-r_g/r)^2 {\vphi_0'}^2 =0, \qquad r=r_s,
\ee
leaving only one term $\propto \dot\vphi_1 \d_\phi \vphi_1$ in \eqref{tauz}. This will greatly simplify the calculation.

%Moreover there is zero torque in $\omega\to 0$ limit. This is manifest in the contribution proportional to $\dot \vphi_1 \d_\phi\vphi_1$, and in the other one follows from the fact that two modes $\vphi_{l_1,m_1}(\omega_1,r)$ and $\vphi_{l_2,m_2}(\omega_2,r)$ have the same $r$-dependence if $l_1=l_2$ and $\omega_1=\omega_2$. Therefore $\vphi_1'\d_\phi \vphi_1$ reduces to a total $\phi$-derivative in this limit. 

Consider a mode with frequency $\omega$ and $l=1$. At linear order $\vphi_{1,m}(\omega,r)$ solves the following equation (we denote $\vphi_{1,m}(\omega,r)$ by $\vphi_1$, and use $u=-(1-r_g/r)\vphi_0'$ to avoid clutter)
\be\label{diff1}
[r^2(1-3u^2)\vphi_1']'-2(1-r_g/r)^{-1}(1-u^2) \vphi_1 = L_\omega \vphi_1,
\ee
where $L_\omega$ is a differential operator that vanishes at $\omega=0$.\footnote{Explicitly
\be
L_\omega = 2i\omega [2r^2 (1-r_g/r)^{-1} u \d_r +((1-r_g/r)^{-1}u r^2)']-\omega^2 r^2(1-r_g/r)^{-2}(3-u^2).
\ee} In the absence of black hole, which implies $r_g=0$ and $u=0$, this equation reduces to $(r^2\vphi_1')'-2\vphi_1=-\omega^2 r^2 \vphi_1$. For $\omega r\ll 1$ there is a growing solution $r$ and a decaying solution $1/r^2$. The asymptotic solution $\vpbg$, given in \eqref{asym}, is a superposition of plane waves and only has the growing component $\r\cdot\nabla\vpbg(\omega,\bsb 0)$. 

In the presence of the black hole $\vphi_1$ matches this asymptotic solution:
\be\label{match}
\sum_{m=-1}^1\vphi_{1,m}(\omega,r\gg r_g)\simeq \r\cdot\nabla\vpbg(\omega,\bsb 0),
%\vphi_{1,0}(\omega,r\gg r_g)\simeq \nabla_z\vpbg(\omega,0) \ r,\qquad 
%\vphi_{1,\pm1}(\omega,r\gg r_g)\simeq(\nabla_x\vpbg(\omega,0)\pm i \nabla_y\vpbg(\omega,0)) \ r.
\ee
with corrections suppressed by $\omega r_g$ or $r_g/r$. These corrections have two sources. (a) The corrections to the growing solution. These can be fixed order-by-order at larger $r$ by demanding that \eqref{diff1} is satisfied. (b) A mixture of the decaying mode due to the sonic singularity of \eqref{diff1} at $r=r_s$ where $1-3 u^2=0$. For a fixed amplitude of the growing mode, there is a unique mixture of the decaying mode such that $\vphi_1$ is regular at $r_s$. In practice, we find this mixture using numerical shooting. 

Let's denote by $\lambda(\omega)$ the ratio of this non-singular solution evaluated at $r_s$ to the coefficient of the growing mode:
\be\label{vphirs}
\sum_{m=-1}^1\vphi_{1,m}(\omega,r_s)=\lambda(\omega) r_g \ \hat r\cdot \nabla \vpbg(\omega,\bsb 0).
\ee
The advantage of evaluating \eqref{tauz} at $r=r_s$ is that, to leading order in $\omega r_g$, the torque can be calculated by solving just for $\lambda(0)$. That is, we can ignore $L_\omega$ in \eqref{diff1} and find the correct mixture of time-independent growing and decaying $l=1$ modes.\footnote{However, with a bit of more work one can calculate the $\O(\omega)$ correction to the mixture and hence \eqref{tauz} at any radius.} We find
\be
\lambda\equiv \lambda(0)\simeq 0.867.
\ee
Next we substitute \eqref{vphirs} in \eqref{tauz} and perform the angular integrals. The result, written in a more covariant way, is
\be\label{Jdot}
\bsb{\dot J}= 2\sqrt{3}\pi \lambda^2 r_g^4 \nabla\rho \times \v.
\ee
Here $\v$ and $\nabla \rho$ are respectively the ambient velocity and density gradient due to the acoustic waves. They are related to the gradients of $\vpbg$ evaluated at $\r =0$ via $\v = \nabla \vpbg$ and $\nabla\rho = -4\bar\rho\nabla\dot \vpbg$.

%%%%%%%%%%%%%%%%%%%%%%%%%%%%%%%%%%%%%%%%%%%%%%%%%%%%%%
\subsection{First order cosmological perturbations}\label{sec:cospert}

We will now relate the perturbations, $\delta \rho$ and $\v$, to the cosmological initial fluctuations $\zeta_\k$. The background parameters in the radiation era and in terms of the conformal time $\eta$ (defined through $ad\eta =dt$) are given by
\be
a = \eta,\qquad \bar \rho = \frac{\rho_0}{\eta^4},\qquad \bar\vphi=-\rho_0^{1/4}\eta,
\ee
where $\rho_0=$constant. The scalar metric perturbations can be parametrized in the conformal Newtonian gauge as
\be
ds^2 = -\eta^2  (1+2\Phi) d\eta^2 + \eta^2 (1-2\Psi) dx^2,
\ee
and the fluid perturbations as
\be
\vphi = \rho_0^{1/4}(-\eta +\hat \vphi).
\ee
So the first order velocity (in the comoving frame) and density perturbations are given by
\be
u_i =\eta \d_i \hat \vphi,\qquad \delta \equiv \frac{\rho-\bar\rho}{\bar\rho}=-4\d_\eta\hat\vphi-4 \Phi.
\ee
Working at finite spatial momentum, the traceless part of the $i-j$ component of Einstein equations gives
\be
\Phi = \Psi.
\ee
The $0-i$ component gives
\be
\hat\vphi = -\frac{1}{2}\eta(\eta\d_\eta\Psi+\Phi).
\ee
The $0-0$ component gives
\be
6\eta \d_\eta \Psi - 2 \eta^2 \delta^{ij}\d_i \d_j \Psi + 3 (\delta +2\Phi) =0.
\ee
From the above three equations we derive an equation for $\Phi$, which after going to the momentum space (with $\bsb k$ the comoving momentum and $k^2=\delta^{ij} k_i k_j$) looks like
\be
\d_\eta^2\Phi+ \frac{4}{\eta}\d_\eta\Phi+ \frac{1}{3} k^2 \Phi =0.
\ee
This has two solutions. The one that is regular at $\eta =0$ is
\be
\Phi_\k(\eta) = 3 \Phi^{(0)}_\k \frac{\sin \omega\eta - \omega \eta \cos \omega \eta}{\omega^3 \eta^3},\qquad 
\omega^2=\frac{1}{3} k^2.
\ee
The initial condition $\Phi_\k^{(0)}$ is related to the primordial fluctuations $\zeta_\k$ by transforming to the comoving gauge \cite{Weinberg} (where up to gradients and time-derivatives the scalar perturbations show up only in the trace of the spatial metric as $\eta^2 (1+2\zeta) dx^2$)
\be
\Phi^{(0)} = -\frac{2}{3} \zeta \qquad \text{Radiation Dominance.}
\ee
We are in particular interested in the sub-horizon ($k\eta \gg 1$) behavior of $\hat\vphi_\k(\eta)$. The gravitational potential is negligible in this limit and we have
\be\label{hvphi}
\hat\vphi_\k(\eta) =  \zeta_\k \frac{\sin (c_s k\eta)}{c_s k}.
\ee
Our torque formula is in terms of the fluid velocity in the physical coordinates $(t,\r=a\x)$, i.e. the coordinates of the inertial frame that surrounds PBH. To evaluate it note that $\bar \rho^{1/4} \vpbg = \rho_0^{1/4}\hat\vphi$ or
\be
\vpbg = \eta \hat\vphi.
\ee
Therefore the physical velocity is given by
\be
v^i=  \frac{\d\vpbg}{d r^i} = \frac{\d\hat \vphi}{\d x^i},
\ee
and our identification $\nabla\rho =-4\bar \rho \nabla\dot\vpbg$ in \eqref{Jdot} receives $1/H$ corrections which are negligible in this regime. Therefore 
\be
\frac{\d\rho}{\d r^i} =-\frac{4\bar \rho}{\eta} \ \frac{\d^2\hat\vphi}{\d\eta\d x^i}.
\ee

%%%%%%%%%%%%%%%%%%%%%%%%%%%%%%%%%%%%%%%%%%%%%%%%%%%%%%
\subsection{RMS angular momentum gain and extrapolation}

The total angular momentum accreted after time $t\gg r_g$ is obtained by integrating the second order torque formula \eqref{Jdot}. $\J$ is defined in an approximately Minkowski frame, so we can safely integrate 
\be
\Delta \J(t) =\int_t^\infty dt_1 \bsb{\dot J}.
\ee
This can be expressed in terms of the cosmological initial conditions (and the conformal time)
\be\label{DJ}
\Delta\J(\eta) =6\pi \lambda^2 r_g^4 \rho_0
\int_\eta^\infty \frac{d\eta_1 }{\eta_1^4}\int_{\k_1,\k_2} \zeta_{\k_1}\zeta_{\k_2}
F_{k_1,k_2}(\eta_1)\ (\hat k_1\times\hat k_2)
\ee
where we symmetrized the expression in $\k_1$ and $\k_2$, and introduced
\be\label{F}
F_{k_1,k_2}(\eta) = k_- \sin(c_s k_+ \eta)-k_+ \sin(c_s k_-\eta),\qquad k_\pm = k_1\pm k_2.
\ee
As seen, unless there are different frequencies in the spectrum, the anti-symmetry of the expression for the torque makes it vanish.

Squaring \eqref{DJ} and taking the expectation value gives $\Delta \Jrms^2(\eta)=\expect{\Delta J(\eta)^2}$
\be\label{J2}
\Delta \Jrms^2(\eta) =
(6\pi \lambda^2 r_g^4 \rho_0)^2
\int_{\eta}^\infty \frac{d\eta_1 d\eta_2 }{\eta_1^4 \eta_2^4}
\int_{\k_1,\k_2} \!\!\! 2 P_\zeta(k_1) P_\zeta(k_2)
F_{k_1,k_2}(\eta_1)F_{k_1,k_2}(\eta_2)(1-(\hat k_1\cdot\hat k_2)^2).
\ee
This can be calculated given a PBH formation scenario, i.e. a specification of $P_\zeta(k)$. Note that because of cosmic expansion the integrals are dominated by the first few Hubble times after $\eta$. Ultimately we let $\eta$ approach the formation period to obtain an estimate of total $\Jrms$. For concreteness, suppose the formation scenario relies on an enhancement of $\zeta$ fluctuations within a relatively narrow $k$-band. Hence the enhanced part of $\zeta$ power spectrum is modeled by a Gaussian
\be
P_\zeta(k)=\frac{2\pi^{3/2}\zrms^2}{k_0^2\Delta k } e^{-{(k-k_0)^2}/{\Delta k^2}},
\ee
where
\be
\zrms^2=\expect{\zeta(\x)\zeta(\bsb 0)}_{x\to 0},
\ee
{\cyan and $\Delta k\ll k_0$. The latter choice serves two purposes. First to demonstrate the extra suppression of $\Jrms$ by $\Delta k/k_0$ as expected from the symmetry argument of section \ref{sec:first}.} Secondly, it allows a further simplification since there is an early period when $k_-\eta\sim \Delta k\ \eta  \ll 1$ but $k_+\eta\sim 2 k_0 \eta\gg 1$. Hence, in the $\eta\to 1/k_0$ limit the dominant contribution to \eqref{J2} comes from the substitution 
\be
F_{k_1,k_2}(\eta_1)\to - k_+ \sin(c_s k_-\eta_1)\simeq -c_s k_+ k_- \eta_1,
\ee
and a similar substitution for $F_{k_1,k_2}(\eta_2)$. The result is
\be
\Delta \Jrms(\eta) \simeq 4\pi \lambda^2 r_g^4 \bar \rho  {\Delta k k_0 \eta^2} \zrms^2.
\ee
One suppression factor in this expression $\Delta k\eta <1$ ensures that for an infinitely narrow spectrum there is no second order torque. The other factor of $k_0\eta>1$ originates from the gradient in the expression for the torque \eqref{Jdot}. {\cyan To express the result in a more useful way denote by $H_0$ the expansion rate at the horizon crossing time of $k_0$. We have $\dot a/a =H_0/(k_0\eta)^2$ and $\bar \rho=3(\dot a/a)^2/8\pi G$, giving
\be
\Delta \Jrms(\eta) \simeq M r_g \zrms^2 \frac{\Delta k}{k_0} \left(\frac{\sqrt{3}\lambda H_0 r_g}{k_0\eta}\right)^2.
\ee
For an average-mass PBH $H_0 r_g\sim 1$. Hence taking the limit $k_0\eta \to 1$ and assuming $\Delta k\sim k_0$, gives the asserted order-1 contribution to the formula \eqref{Jrms}.}

%%%%%%%%%%%%%%%%%%%%%%%%%%%%
\section{Conclusions}\label{sec:con}

We explained in what sense the spin of PBHs is small, arguing that 
\be
\label{Jrms2}
\Jrms\sim M r_g \zrms^2.
\ee
This was supported by calculating the torque due to the acoustic waves long after black hole formation and by extrapolating the result to early times. We also showed that for a narrow enhancement of the primordial power $P_\zeta(k)$ the second order $\Jrms$ is suppressed by the width $\Delta k/k_0$. 

The early contribution to $J$ during the formation period seems beyond analytic reach. However, given that it has to vanish at zeroth and first order in $\zrms$, it is natural to expect it to have a similar size as \eqref{Jrms2}. Although a partial cancellation is in principle possible, a full cancellation seems unlikely. Clearly, at late enough times the torque (though small) is uncorrelated with the formation process because of the propagation of the sound waves. 

For PBHs to form a sizable fraction of dark matter, the primordial fluctuations cannot be much smaller than $\zrms\sim 0.1$ as long as the Gaussian approximation is valid. The resulting spin parameter $a\sim 0.01$ is small, but it is not unimaginable that such a precision is reached in the future.

%%%%%%%%%%%%%%%%%%%%%%%%%%%%%
%%%%%%%%%%%%%%%%%%%%%%%%%%%%%%
%%%%%%%%%%%%%%%%%%%%%%%%%%%%
%%%%%%%%%%%%%%%%%%%%%%%%%%%
%%%%%%%%%%%%%%%%%%%%%%%%%%

\vspace{0.3cm}
\noindent
\section*{Acknowledgments}

We thank Kfir Blum for posing the question to us and Diptimoy Ghosh, Francesco Muia, Yacine Ali-Haimoud, Marcello Musso, and Ravi Sheth for useful discussions. The work of MM was partially supported by the Simons Foundation Origins of the Universe program (Modern Inflationary Cosmology collaboration).

\bibliography{bibspbh}
\end{document}